\documentclass[aps,prl,reprint,showpacs,groupedaddress]{revtex4-1}

\usepackage[utf8]{inputenc}
\usepackage[T1]{fontenc}
\usepackage{amsmath}
\usepackage{amsfonts}
\usepackage{amssymb}
\usepackage{siunitx}
\usepackage[colorlinks]{hyperref}
\hypersetup{pdfdisplaydoctitle=true, colorlinks=true, linktocpage=false,  linkcolor=black, menucolor=black,
   urlcolor=black, citecolor=black, bookmarksopen=false, bookmarksnumbered=true
}
\usepackage[pdftex]{graphicx}
\graphicspath{{figures/}}

\begin{document}

\title{Motional Coherence of Fermions Immersed in a Bose Gas}

\author{R. Scelle}
\email[]{decoherence@matterwave.de}
\author{T. Rentrop}
\author{A. Trautmann}
\author{T. Schuster}
\author{M. K. Oberthaler}
\affiliation{Kirchhoff Institute for Physics, University of Heidelberg, INF 227, 69120 Heidelberg, Germany}

\bibliographystyle{apsrev4-1}

\date{\today}

\begin{abstract}
We prepare a superposition of two motional states by addressing lithium atoms immersed in a Bose-Einstein condensate of sodium with a species-selective potential. The evolution of the superposition state is characterized by the populations of the constituent states as well as their coherence. The latter we extract employing a novel scheme analogous to the spin-echo technique. Comparing the results directly to measurements on freely-evolving fermions allows us to isolate the decoherence effects induced by the bath. In our system, the decoherence time is close to the maximal possible value since the decoherence is dominated by population relaxation processes. The measured data are in good agreement with a theoretical model based on Fermi's golden rule.
\end{abstract}


\maketitle

Dissipation due to coupling of a quantum system to a bath is of fundamental interest since it provides a route from quantum to classical dynamics via decoherence~\cite{Zurek03,Schlosshauer05}. Recently, it has been suggested that dissipative processes also provide access to a new class of many particle systems building on the possibilities offered by mixtures of ultracold gases~\cite{Zoller08}. 

Decoherence of immersed quantum systems~\cite{Leggett87} has been experimentally studied in a variety of systems such as atoms and ions exposed to electric as well as light fields~\cite{Mlynek94,Haroche96,Ertmer00,Wineland00,Blatt04,Wineland05,Blatt11}, and spins interacting with a spin bath~\cite{Awschalom08,Koehl13}. The demonstration and systematic study of motional decoherence with fermions immersed into a Bose Einstein condensate is reported here. This new experimental model system opens the route for further studies in the direction of many particle phenomena such as the emergences of Fröhlich polarons~\cite{Timmermans06,Jaksch07,Tempere09}, and also energy dissipation dynamics for supersonic motion of impurities moving in a superfluid~\cite{Kovrizhin01,Komnik13}. The detailed study of the coherence loss also puts an important limit on the applicability of oscillation measurements for inferring the effective mass of impurities~\cite{Inguscio2012}. 

\begin{figure}[b]
    \centering
    \includegraphics[width=86mm]{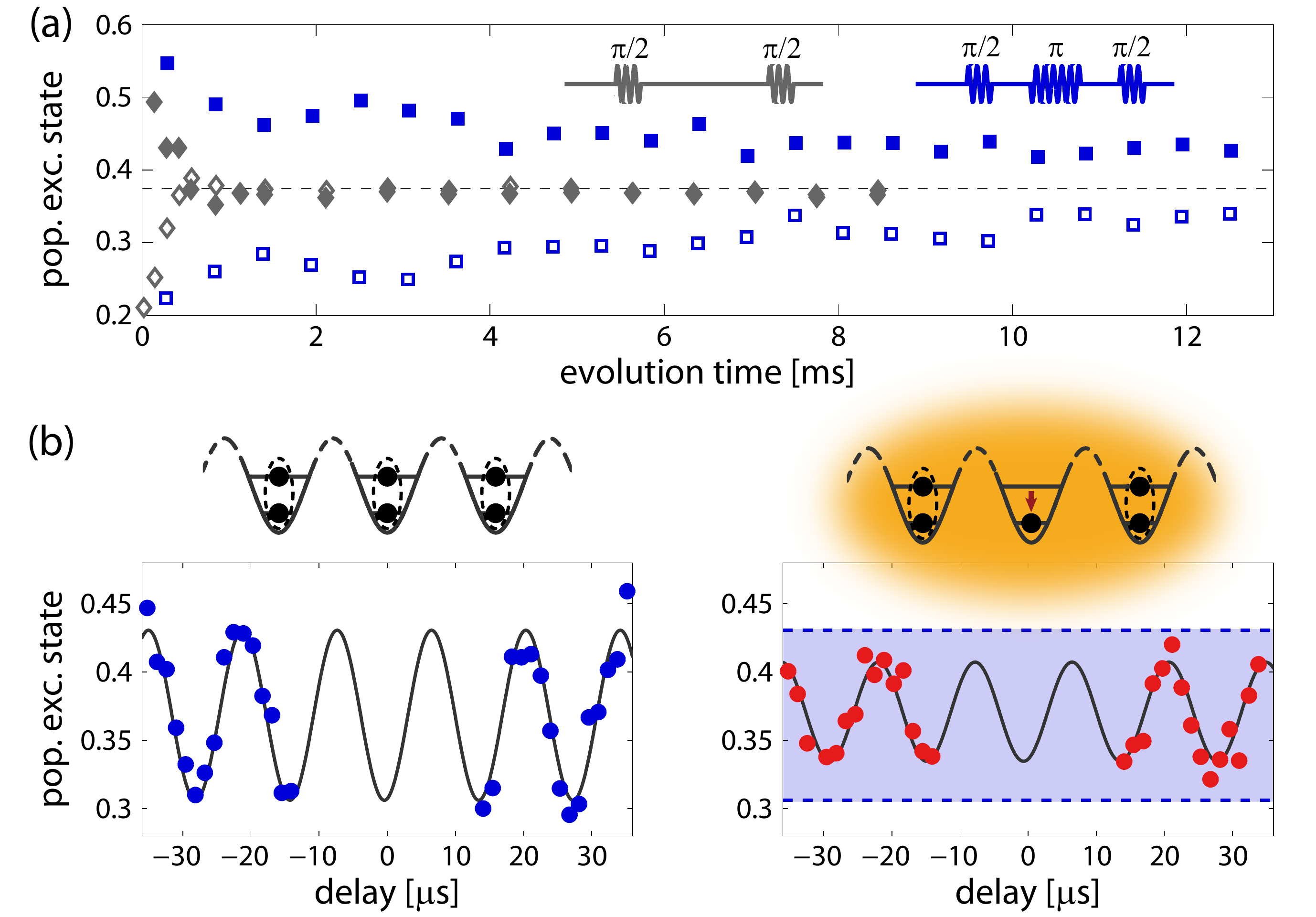}
    \caption{Detection of motional coherence with and without bath.  (a)  Shaking the position of the optical 
	lattice as indicated in the insets implements motional Ramsey (gray) and spin-echo spectroscopy (blue).	
	Motional coherence is inferred from the interference amplitude which is determined by the population in the 
	excited state after evolution times corresponding to zero/$\pi$ phase of the last Ramsey pulse. The results 
	are depicted as solid/open symbols. (b) Motional interference fringes for the spin-echo sequence obtained by 
	a short delay of 	the second $\pi/2$-pulse. The signal without/with bath (present for \SI{2.8}{ms}) is shown 
	in the left/right panel. The blue shading in the right panel indicates the signal height without bath and clearly 
	reveals decoherence.}
    \label{fig_intro}
\end{figure}

In our experiment, fermionic $^{6}$Li is confined in a deep species-selective one-dimensional lattice potential and is immersed in a Bose-Einstein condensate of $^{23}$Na. The lattice parameters are chosen such that tunneling is negligible and thus the dynamics along the lattice axis, here referred to as longitudinal direction, can be described by a set of independent harmonic oscillators. For the manipulation and preparation of the motional quantum states of $^{6}$Li, we periodically modulate the lattice position which coherently couples the two lowest longitudinal states~\cite{Philipps2002} which are in the following denoted as ground and excited state. To extract the motional coherence, we realize a Ramsey sequence as indicated in the left inset of Fig.\;\ref{fig_intro}(a). This is implemented by shaking the lattice potential for a specific time ($\pi/2$-pulse) preparing the $^{6}$Li atoms in an equal superposition of ground and excited state. After a subsequent free evolution, we apply a second $\pi/2$-pulse with variable delay and infer the remaining motional coherence from the amplitude of the interference signal deduced from the population in the ground and excited state. The envelope of the interference signal shown in Fig.\;\ref{fig_intro}(a) (gray diamonds) reveals a fast decay in \SI{1}{ms}. This decay results primarily from dephasing due to the inhomogeneous distribution of trapping frequencies. The implementation of a spin-echo sequence, by introducing a rephasing $\pi$-pulse, compensates the dephasing effects and enables the observation of the interference signal for more than \SI{11}{ms} of free evolution corresponding to \num{800} harmonic oscillation cycles (blue squares in Fig.\;\ref{fig_intro}(a)). 
These long observation times provide the access to the regime of decoherence that is intrinsically slow, due to the weak coupling to the bosonic bath. In Fig.\;\ref{fig_intro}(b) the interference signals without and with bath (present for \SI{2.8}{ms}) are compared which reveal  the impact of a three-dimensional degenerate Bose gas on the motional coherence. We observe a clear decay of the oscillation amplitude, i.\,e.~decoherence between the ground and excited state.

Before presenting the results of our detailed studies, we introduce our experimental setup. We produce a mixture of bosonic $^{23}$Na and fermionic $^{6}$Li in a crossed beam optical dipole trap ($\lambda=1064\,$nm) with a geometric mean trapping frequency of $\bar{\omega} =  2 \pi \! \cdot  \! \SI{238}{Hz}$ for $^{23}$Na and $2 \pi \! \cdot  \! \SI{500}{Hz}$ for $^{6}$Li. Because of the large trapping frequencies and the attractive interaction between bosons and fermions, the differential gravitational sag is about a factor \num{10} smaller than the size of the $^{6}$Li cloud and is hence negligible for our experiments. Both species are prepared in the respective hyperfine ground state. The  one-dimensional species-selective lattice for $^{6}$Li~\cite{LeBlanc07} is created by two laser beams ($\lambda=670.5\,\text{nm}$, \SI{550}{\micro m} waist) intersecting at 35$^\circ$ which results in a lattice spacing of $d_{\text{lat}}=\SI{1.1}{\micro m}$ and a corresponding recoil energy for $^{6}$Li of $E_{\text{r}}=h^{2}/(8 m_{\text{Li}} d_{\text{lat}}^{2})=h \! \cdot \!  \SI{6.9}{kHz}$. The residual potential for $^{23}$Na is more than \num{100}-times lower and thus negligible. For preparing the fermions in harmonic traps, we exponentially increase the lattice depth within \SI{100}{ms} to $V_{\text{Li}}=$ \SI{33}{$E$_{r}} yielding an energy gap of  $\Delta E=$ \SI{11}{$E$_{r}} between the ground and excited state. We determine the populations of these states by a band mapping technique~\cite{Esslinger01}, which is achieved by ramping down the lattice potential exponentially within \SI{500}{\micro s} and imaging the $^{6}$Li cloud after \SI{4.5}{ms} time-of-flight. For our typical atom numbers ($N_{\text{Li}}=2 \cdot 10^{5}$, $N_{\text{Na}}=5 \cdot 10^{5}$) and temperature ($T \approx \SI{600}{nK}$), \SI{85}{\%} of the $^{6}$Li atoms populate the first Bloch band, i.\,e.~the ground state. It is interesting to note that ramping up the lattice without $^{23}$Na background, we observe an occupation of the first Bloch band of \SI{70}{\%} due to the absence of relaxation processes~\cite{Smerzi04}. Thus, we attribute the remaining population in the excited state in the presence of the $^{23}$Na atoms mainly to the vanishing relaxation rates for the  $^{6}$Li atoms not overlapping with the $^{23}$Na cloud. After the preparation, our measurements start typically with a $^{6}$Li cloud having a peak density of \SI{1.2e13}{cm^{-3}} immersed in $^{23}$Na  Bose-Einstein condensate with a condensate fraction of \SI{60}{\%} and a peak density of \SI{3.7e14}{cm^{-3}}. For the generation and detection of the motional coherence by the $\pi$- and $\pi/2$-pulses, the lattice position is modulated by \num{0.01} lattice spacings which corresponds to a Rabi frequency of $\Omega =  2 \pi \! \cdot  \! \SI{3.5}{kHz}$. 

Investigating the dynamics in an ultracold gas gives us the opportunity to directly compare  a freely evolving system and a system immersed in a bath. Therefore, we can isolate the impact of the bosonic bath and distinguish between interaction induced decoherence and other mechanisms for coherence loss, e.\,g.~dephasing due to technical imperfections and residual tunneling between neighboring lattice sites. Tunneling dynamics and excitations in the transverse directions cause the observed oscillatory behavior of the spin-echo signal shown in Fig.\;\ref{fig_intro}(a). The influence of tunneling processes has been confirmed in separate experiments by increasing the harmonic confinement along the lattice axis leading to the suppression of tunneling (Wannier-Stark effect). By adding lattice potentials in each of the transverse directions, we observe a reduction of the amplitude of the oscillatory features. Since these species-selective potentials rely on near-resonant light, the available observation time is for the three-dimensional lattice reduced due to residual spontaneous emission. Therefore, we use, for the following study, a one-dimensional lattice, i.\,e.~two-dimensional Fermi gases immersed in a Bose gas.

\begin{figure}[t]
    \centering
    \includegraphics[width=86mm]{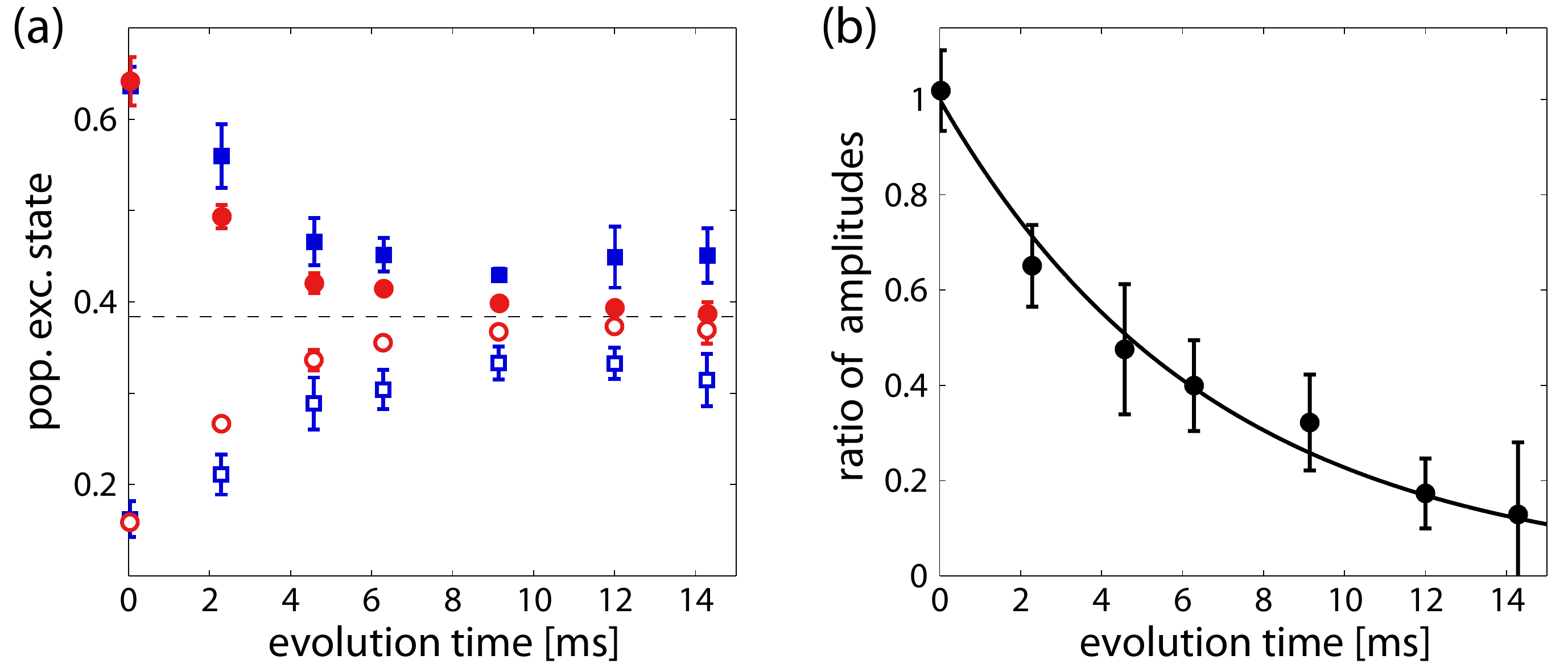}
    \caption{Systematic measurement of the decoherence time $\text{T}_{\text{2}}$. (a) The interference 
	amplitudes for zero/$\pi$ phase (solid/open symbols)  reveal the faster signal decay in presence of the bath 
	(circles) compared to no bath (squares). Because of the different sampling, the spin-echo signal does not 
	show the oscillatory behavior apparent in Fig.\;\ref{fig_intro}(a), but only one revival at long evolution times.  
	(b) The decoherence time $\text{T}_{\text{2}}=\SI{6.8\pm0.4}{ms}$ is deduced from 
	the exponential fit to the ratio of the interference amplitudes with and without bath.}
    \label{Fig_dec}
\end{figure}

In order to determine the decoherence time, referred to as  $\text{T}_{\text{2}}$ in the context of nuclear magnetic resonance~\cite{Levitt2002}, we compare the evolution of the $^{6}$Li atoms with and without bath. Fig.\;\ref{Fig_dec}(a) summarizes our observations revealing that after \SI{12}{ms} the coherence is lost in presence of the Bose gas. By analyzing the ratio between the fringe amplitudes with and without bath as shown in  Fig.\;\ref{Fig_dec}(b), $\text{T}_{\text{2}}$ can be extracted independent of a residual time-dependence of the spin-echo signal. By fitting with an exponential, we infer the characteristic decay time of $\text{T}_{\text{2}}=\SI{6.8\pm0.4}{ms}$ for the $^{6}$Li cloud experiencing an averaged sodium density of  $\bar{n}_{\text{Na}}=\int \! n_{\text{Na}}(r) n_{\text{Li}}(r)\text{d}V/N_{\text{Li}} = \SI{1.8e13}{cm^{-3}}$, where $n_{\text{Na}}(r)$ and $n_{\text{Li}}(r)$ denote the sodium and lithium density, respectively. All indicated errors are statistical errors and denote one standard deviation.

\begin{figure}[tb]
    \centering
    \includegraphics[width=86mm]{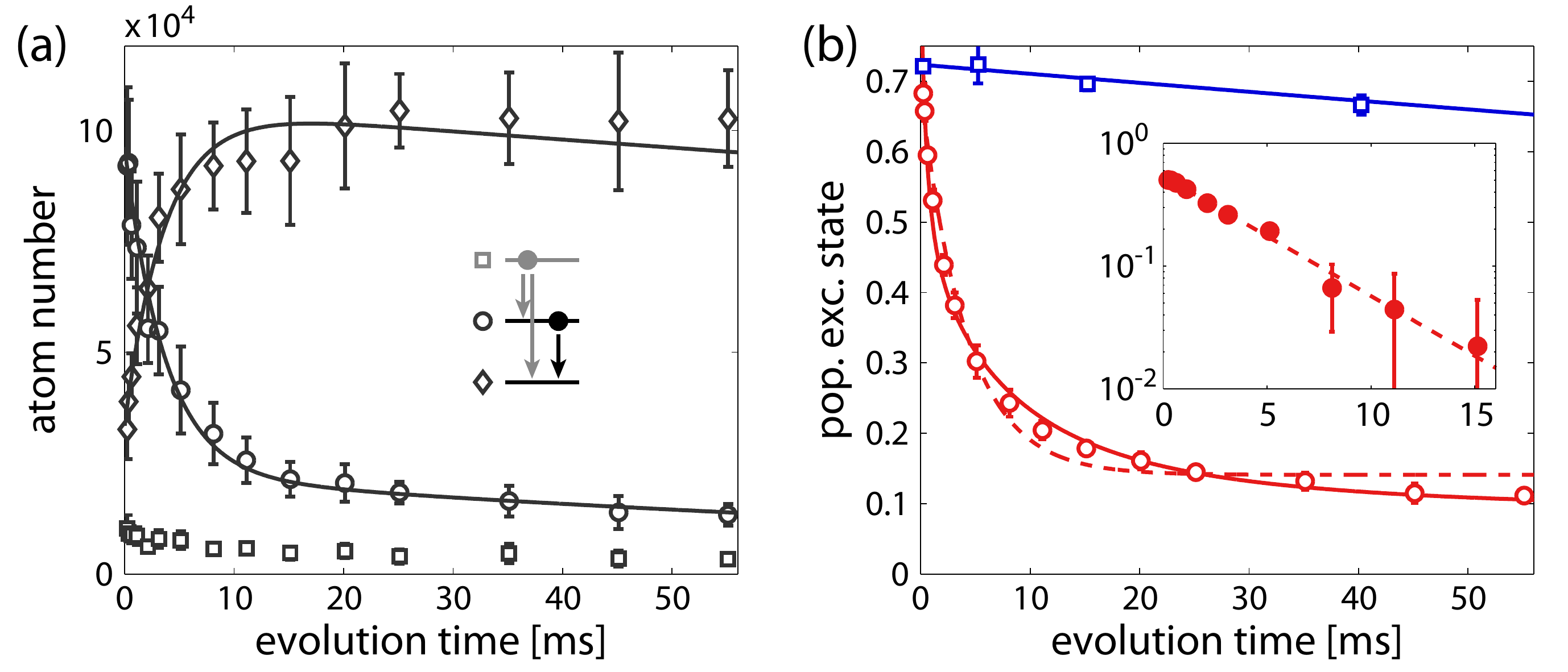}
    \caption{Systematic investigation of the relaxation time  $\text{T}_{\text{1}}$. (a) The $^{6}$Li atom numbers 
	in the ground (diamonds), first excited (circles) and second excited (squares) state reveal the 
	relaxation to the ground state. The third level is occupied by a negligible number of atoms
	and does not contribute to the dynamics. (b) For long evolution times, the normalized population in the 
        first excited state (open circles) deviates from an exponential fit (dashed line). Including the inhomogeneous 
        density distribution of the $^{6}$Li and $^{23}$Na atoms (solid line) explains the deviation. The population in 
	the excited state without bath (squares) is consistent with the expected loss due to spontaneous emission of 
	the species-selective potential. The inset reveals that the population of the excited state decays for short times 	exponentially. For clarity, we subtract the observed long term background due to inhomogeneous density 
	distributions. The dashed line shows an exponential fit with a relaxation time  	
	$\text{T}_{\text{1}}=\SI{4.5\pm0.4}{ms}$.}
    \label{Fig_relax}
\end{figure}

More insight into the microscopic processes leading to decoherence is gained by investigating the population decay of the excited state, i.\,e.~energy relaxation dynamics (see e.\,g.~\cite{Widera2012}). For this purpose, we prepare the atoms with a $\pi$-pulse in the first excited state and study the evolution of the populations of the three lowest states extracted from the band mapping images (Fig.\;\ref{Fig_relax}(a)). The small atom number in the second excited state (third band) supports the simplified description of our system as two level system. The initial time evolution clearly reveals the relaxation dynamics as the $^{6}$Li atoms in the first excited state decay mainly into the ground state. The solid lines represent the result of a rate equation model additionally taking into account the minor effect of the independently determined loss due to spontaneous emission. The model captures the physics for the time scales of interest ($\text{T}_{\text{2}}$), and implies an exponential population decay.
For a quantitative determination of the relaxation time $\text{T}_{\text{1}}$, we analyze the population in the excited state normalized to the total number of particles in order to reduce the statistical errors originating from the shot-to-shot fluctuations of the $^{6}$Li atom number. The dashed line in Fig.\;\ref{Fig_relax}(b) represents an exponential fit to the data which gives good agreement for short times. To extract $\text{T}_{\text{1}}$, we analyze the first \SI{15}{ms} as indicated in the inset and obtain $\text{T}_{\text{1}}=\SI{4.5\pm0.4}{ms}$. The solid line in Fig.\;\ref{Fig_relax}(b) showing better agreement over the whole evolution results from a model accounting for the inhomogeneous sodium and lithium density distributions. For this purpose, we sum over the exponential decays of the different lattice sites assuming a linear dependence of the relaxation rate with the sodium density $n_{\text{Na}}(r)$.  

For a quantitative comparison of experimentally observed decay time and theoretical predictions, we estimate the energy relaxation rates $R^{-}_{n \rightarrow m}$ according to  Fermi's golden rule~\cite{Zoller04} 
\begin{equation}
 R^{-}_{n \rightarrow m}  =   \frac{2 \pi}{\hbar} \sum_{\vec{q}} \left(1\!+\!N_{\vec{q}}\right) \left| V_{m,n}(\vec{q})  \right|^{2}
 \delta(E_{n}\!-\!E_{m}\!-\!\epsilon_{\vec{q}}),
  \label{eqRelRate}
\end{equation}
where $n$ and $m$ refer to the initial and final three-dimensional harmonic oscillator state with the eigenenergies $E_{\text{n}}$ and $E_{\text{m}}$ (the superscript minus indicates $E_{\text{n}} > E_{\text{m}}$). $N_{\vec{q}}$ denotes the initial occupation of the Bogoliubov excitations according to the Bose distribution. These excitations are characterized by the quasimomentum $\vec{q}$ and the energy  $\epsilon_{\vec{q}} = \sqrt{(\hbar \vec{q})^{2}\mu/m_{\text{Na}} + ((\hbar \vec{q})^{2}/2m_{\text{Na}})^{2}}$ with $\mu$ being the chemical potential of the Bose-Einstein condensate. $V_{m,n}(\vec{q}) = \langle m | \hat{V}(\vec{q}) | n \rangle$ denotes the matrix element for the initial and final state accounting for the interaction between the lithium and sodium atoms 
\begin{equation}
  \hat{V}(\vec{q}) = (u_{\vec{q}}+v_{\vec{q}}) \, g_{\text{NaLi}} \, \sqrt\frac{n_{\text{Na}}}{V_{\text{N}}} e^{-i \vec{q} \, \hat{\vec{r}}}
\end{equation}
with $V_{\text{N}}$ being the normalization volume. $g_{\text{NaLi}} =  2 \pi \hbar^{2} a_{\text{NaLi}} (m_{\text{Na}}+m_{\text{Li}})/ m_{\text{Na}} m_{\text{Li}}$ refers to the coupling constant between the two species with the scattering length $a_{\text{NaLi}}\approx \SI{-75}{a_{0}}$~\cite{Tiemann12}, $\text{a}_{\text{0}}$ being the Bohr radius. $u_{\vec{q}}=\sqrt{ (\hbar^{2} \vec{q}^2/2m_{\text{Na}}+\mu)/2\epsilon_{\vec{q}}+1/2}$ and $v_{\vec{q}}=-\sqrt{ (\hbar^{2} \vec{q}^2/2m_{\text{Na}}+\mu)/2\epsilon_{\vec{q}}-1/2}$ are the coefficients of the Bogoliubov transformation in the homogeneous situation.

For our experimental situation, the energy gap between the ground and excited state ($\Delta E\approx \SI{70}{kHz}$) is much larger than the chemical potential of the $^{23}$Na atoms ($\mu \approx \SI{6}{kHz}$). Thus, the relaxation processes involve, to a good approximation, only real particle excitations of the Bose gas. This has been confirmed by numerically solving equation\;(\ref{eqRelRate}). The corresponding matrix elements $V_{m_{\text{g}},n_{\text{e}}}(\vec{q})$ reflect that real particle excitations dominate the relaxation process for the experimentally relevant transverse excitations of the ground state $m_{\text{g}}$ and excited state $n_{\text{e}}$. The relaxation times  $1/\text{T}_{\text{1}} = \sum_{m_{\text{g}}} R^{-}_{n_{\text{e}} \rightarrow m_{\text{g}}}$ to the ground state manifold show, for the different initial transverse excitations  $n_{\text{e}}$, only a weak dependence for our experimental parameters (less than \SI{5}{\%}). For a quantitative comparison with the experimentally observed decay time, we take the mean relaxation time and build on the linear scaling of the relaxation rate with the sodium density in the case of real particle excitations. With that we can account for the inhomogeneity of the sodium background by the averaged sodium density $\bar{n}_{\text{Na}}=\SI{2.0e13}{cm^{-3}}$ and obtain $\text{T}_{\text{1}} \approx \SI{4.6}{ms}$ (without free parameters) being in good agreement with our observation of $\text{T}_{\text{1}}=\SI{4.5\pm0.4}{ms}$. It is interesting to note that the estimate of the classical energy relaxation time $\text{T}_{\text{1}} = 1/(\sigma \bar{n}_{\text{Na}} v)$ with the cross section $\sigma = 4 \pi a_{\text{NaLi}}^{2}$  and the additional assumption that the relevant velocity can be estimated  from the momentum uncertainty of the excited state ($v = \sqrt{ \langle v^{2} \rangle}$) leads to  $\text{T}_{\text{1}} \approx \SI{3.0}{ms}$.

\begin{figure}[tb]
    \centering
    \includegraphics[width=86mm]{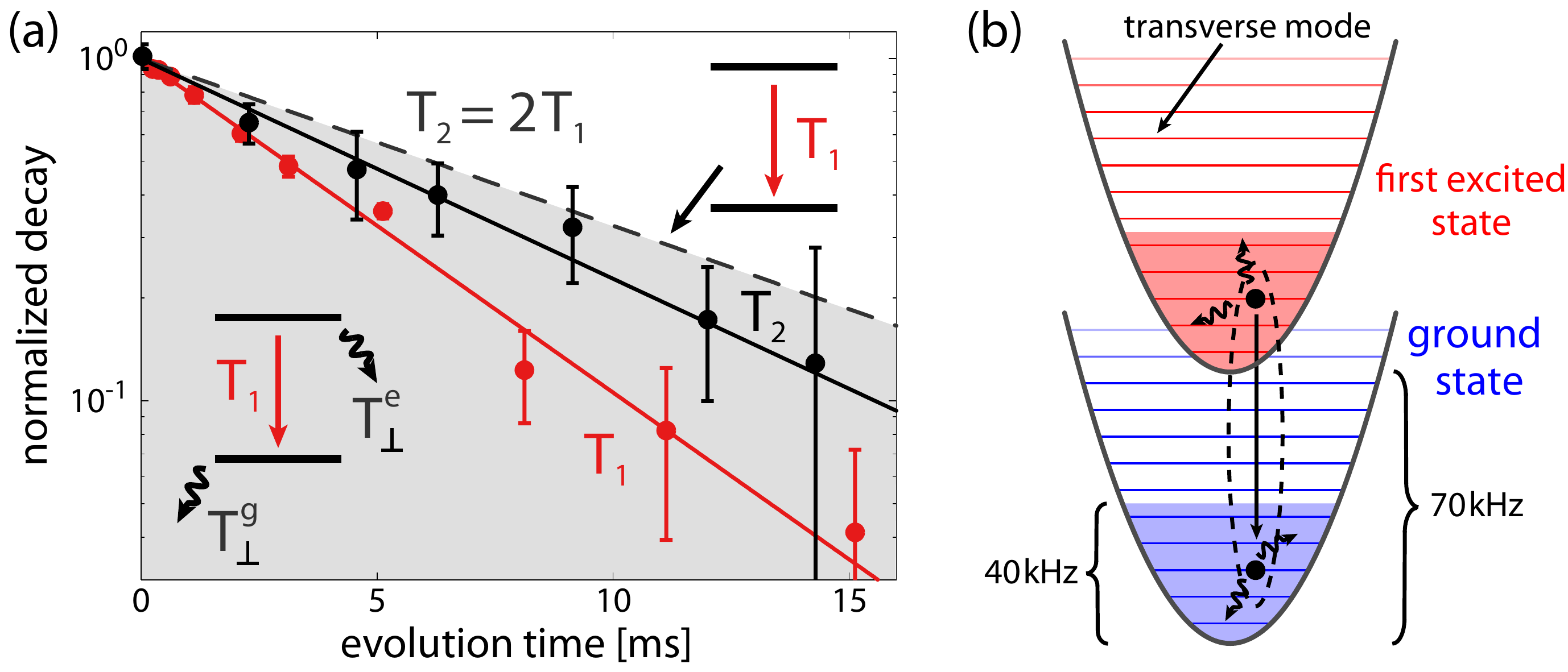}
    \caption{(a) Connection between relaxation and decoherence time. For a two-level system coupled to a 
	Markovian 	bath $\text{T}_{\text{2}}= 2 \, \text{T}_{\text{1}}$ as indicated with the dashed line. Additional 
	coupling of the states to other unobserved degrees of freedom, represented by the wiggly arrows, 
	leads to a reduction of the ratio to $\text{T}_{\text{2}}=(1.5 \pm 0.1) \, \text{T}_{\text{1}} $. (b) 
	Schematic representation of the microscopic processes leading  to $\text{T}^{\text{g,e}}_{\bot}$ involve 
	phonon-like, low energetic elementary excitations of the bath. The shadings display the chemical 
	potential of $^{6}$Li  which indicates the occupied transverse modes.}
    \label{Fig_comp}
\end{figure}

In Fig.\;\ref{Fig_comp}(a) we compare the observed population and coherence decay. For a closed two-level system at zero temperature, the decoherence  time is given by $\text{T}_{\text{2}}= 2 \, \text{T}_{\text{1}}$ as indicated by the dashed line. Experimentally we observe that $\text{T}_{\text{2}} = (1.5 \pm 0.1) \, \text{T}_{\text{1}}$ revealing that other decoherence processes are also present in our system. However, as $\text{T}_{\text{2}}> \text{T}_{\text{1}}$, the relaxation processes from the excited to the ground state can be regarded as the dominant decoherence mechanism. We attribute the additional coherence decay to coupling to transverse excitations within the ground and excited  manifold respectively, as indicated by the wiggly arrows in the inset in Fig.\;\ref{Fig_comp}(a) and (b). This interpretation is confirmed by our observation of  $\text{T}_{\text{2}} = (2.1 \pm 0.3)\, \text{T}_{\text{1}}$ in a deep three-dimensional lattice potential for which transverse processes are negligible. 
 
More insight into the role of the transverse degrees of freedom for decoherence can be gained by assuming uncorrelated loss processes for the ground and excited state, e.\,g.~heating and relaxation to other transverse states. We quantify these processes by the characteristic time scales $\text{T}^{\text{g}}_{\bot}$ and $\text{T}^{\text{e}}_{\bot}$, respectively. In this case the decoherence time is given by
\begin{equation}
  \frac{1}{\text{T}_{\text{2}}} =  \frac{1}{2 \text{T}_{\text{1}}} +\frac{1}{2 \text{T}^{\text{g}}_{\bot}}+\frac{1}{2 \text{T}^{\text{e}}_{\bot}}.
\end{equation}
We estimate $1/\text{T}^{\alpha}_{\bot} = \sum_{m_{\alpha}} \left( R^{-}_{n_{\alpha} \rightarrow m_{\alpha}} + R^{+}_{n_{\alpha} \rightarrow m_{\alpha}}  \right)$  on the basis of the relaxation rates  $R^{-}_{n \rightarrow m}$ and 
heating rates  $R^{+}_{n \rightarrow m}$~\cite{Zoller04} 
\begin{equation}
 R^{+}_{n \rightarrow m}  =   \frac{2 \pi}{\hbar} \sum_{\vec{q}} N_{\vec{q}}  \left| V_{m,n}(\vec{q})  \right|^{2}
 \delta(E_{m}\!-\!E_{n}\!-\!\epsilon_{\vec{q}}),
  \label{eqHeatRate}
\end{equation}
where the superscript plus indicates $E_{\text{n}} < E_{\text{m}}$. The initial assumption of uncorrelated loss processes implies that the scattering processes carries away information about the constituents of the coherent superposition~\cite{Pritchard1995}. This is not true for all transverse relaxation and heating processes which involve low-energetic and thus long-wavelength excitations. These processes do not inevitably cause the loss of coherence since the details of the longitudinal wavefunctions are not resolved. This phenomenon has been explicitly observed for ions exposed to a light field~\cite{Wineland05} or molecules immersed in a solvent~\cite{Tokmakoff04}. In order to take this effect phenomenologically into account, we limit the sum in eq.\;(\ref{eqRelRate}) and (\ref{eqHeatRate}) to the momenta with longitudinal components $q_{\parallel}>\sqrt{m_{\text{Li}} \Delta E}/\hbar$, where the right hand side is the inverse of the harmonic oscillator length of the lattice potential. Assuming an initial superposition with identical transverse excitation as indicated in Fig.\;\ref{Fig_comp}(b), we obtain $\text{T}_{\text{2}} < 1.7 \, \text{T}_{\text{1}}$ for the densities realized in the $^{23}$Na cloud. This is consistent with our observation of $\text{T}_{\text{2}} = (1.5 \pm 0.1) \, \text{T}_{\text{1}}$.

In conclusion, we utilize fermions in a species-selective potential immersed in a Bose gas to investigate decoherence and relaxation dynamics in a well-controlled system. For this purpose, we have developed a new tool for investigating coherence of external degrees of freedom, namely motional spin-echo spectroscopy. Our experiments show that the decoherence time is longer than the relaxation time implying that the two level system can be regarded as well isolated from other motional degrees of freedom.  Furthermore, the experimental technique of motional spin-echo spectroscopy developed here opens new possibilities in studies of motional coherence and precise oscillation frequency measurements. We also anticipate that further refinements of this method will allow us to resolve small changes in effective masses due to the coupling to a background~\cite{Timmermans06,Jaksch07,Tempere09}.

\begin{acknowledgments}  
  We thank Isabelle Bouchoule for enlightening discussions in the context of decoherence phenomena, and
  Andreas Komnik, and J\"org Evers for discussions on the dynamics of impurities immersed in a Bose gas. 
  This work was supported by the Heidelberg Center of Quantum Dynamics, the ExtreMe Matter Institute, and 
   the FET-Open project QIBEC (Contract No.~284584).
  We acknowledge support from the Heidelberg Graduate School of Fundamental Physics, 
   and R.~S.~from the Landesgraduiertenf\"orderung Baden-W\"urttemberg.
\end{acknowledgments}

\end{document}